\documentclass[doublecol]{epl2}
 \usepackage{amsmath}
\title{Wave turbulence in a two-layer fluid: coupling between\\ free surface and interface waves}
\shorttitle{Wave turbulence at the interface between two fluids} %Insert here a short version of the title if it exceeds 70 characters

\author{Bruno Issenmann\inst{1,2}, Claude Laroche\inst{1} \and Eric Falcon\inst{1}}
\shortauthor{B. Issenmann \etal}

\institute{                    
  \inst{1} Universit\'e Paris Diderot, Sorbonne Paris Cit\'e, MSC, UMR 7057 CNRS - F-75013 Paris, France\\
  \inst{2} Univ Lyon, Universit\'{e} Claude Bernard Lyon 1, CNRS, Institut Lumi\`{e}re Mati\`{e}re, F-69622, Villeurbanne, France
}

\abstract{We experimentally study gravity-capillary wave turbulence on the interface between two immiscible fluids of close density with free upper surface. We locally measure the wave height at the interface between both fluids by means of a highly sensitive laser Doppler vibrometer. We show that the inertial range of the capillary wave turbulence regime is significantly extended when the upper fluid depth is increased: The crossover frequency between the gravity and capillary wave turbulence regimes is found to decrease whereas the dissipative cut-off frequency of the spectrum is found to increase. We explain these observations by the progressive decoupling between waves propagating at the interface and the ones at the free surface, using the full dispersion relation of gravity-capillary waves in a two-layer fluid of finite depths. The cut-off evolution is due to the disappearance of parasitic capillaries responsible for the main wave dissipation for a single fluid.}

\begin{document}

\maketitle

\section{Introduction}
Stratified fluids are ubiquitous in Nature such as in ocean or in atmosphere. The density stratification is usually due to a temperature or salinity gradient with the depth in oceans, or a temperature or humidity gradient with altitude in the atmosphere. The simplest stratified fluid consists in two superimposed homogeneous fluids, the fluid with higher density being below the fluid with lower density. In this situation, waves can propagate at the interface between the two fluid layers but also at the free surface of the top one. Under certain conditions, surface and interface waves interact together \cite{Mohapatra2011,Jamali2003}. An astonishing manifestation of this phenomenon is the dead-water effect first observed in 1904 on the sea surface \cite{Ekman1904}, and recently reproduced in experiments \cite{Mercier2011,Grue2015}. Indeed, ships evolving on a calm sea can slow or even stop sailing in a two-layer fluid due to the extra-drag generated by large interface waves. The coupling between the surface and interface waves in a two-layer fluid also generates narrow nested V-shaped wakes observed behind ships \cite{Stapleton1995,Wei2005}, as well as the damping of ocean surface waves over a layer of fluid mud \cite{Trowbridge2015}. Such a coupling is also involved in Faraday instability of floating droplets on a liquid bath \cite{Pucci2013,Pototsky2016}, or during the long-wave instabilities in thin two-layer liquid films ($< 100$ nm) in chemical physics \cite{Pototsky2005}. In industrial applications like metal refining, such interactions can also have an influence on the ripples created during dewetting \cite{Peron2012}. At last, the coupling between the surface and interface waves of large amplitudes occur in many physical and biological situations (involving or not elasticity), and lead to numerous challenging studies in applied mathematics such as the predictions of new solitary waves \cite{Woolfenden2011,Korobkin2011}.

When a set of stochastic waves, propagating on a free surface, have large enough amplitudes, interactions between nonlinear waves can generate a wave turbulence regime. These interactions transfer the wave energy from the large scales, where it is injected, to the small scales where it is dissipated. This generic phenomenon concerns various domains at different scales: Surface and internal waves in oceans, elastic waves on plates, spin waves in solids, magnetohydrodynamic waves in astrophysical plasma (for reviews, see~\cite{Falcon2010,Zakharovbook,Nazarenkobook,Newell2011}). Weak turbulence theory developed in the 60's\cite{Hasselmann1962,Benney1967,Zakharov1967} leads to predictions on the wave turbulence regime in almost all domains of physics involving waves \cite{Zakharovbook,Nazarenkobook}. The past decade has seen an important experimental effort to test the validity domain of weak turbulence theory on different wave systems (e.g. hydrodynamics, optics, hydro-elastic or elastic waves) \cite{NazarenkoAdvance2013}.
%\cite{Falcon2007,Mordant2010,Cobelli2011,Kuznetsov1991}.

\begin{figure}[t]
\onefigure[width=8cm]{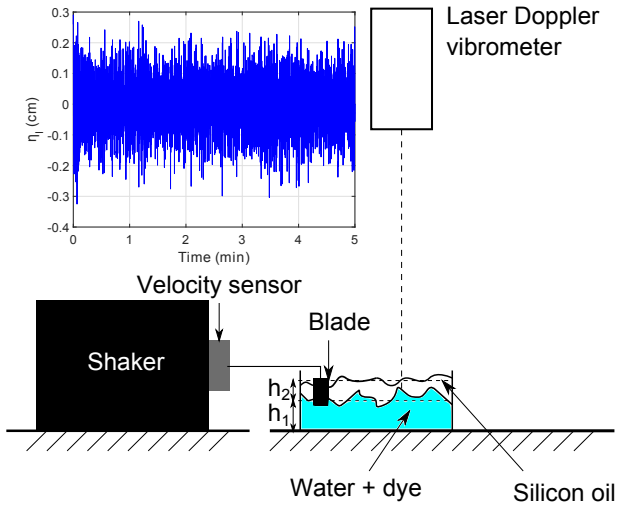}
\caption{Experimental setup. Free surface and interfacial waves are generated by a wavemaker. Lower fluid: water (depth at rest $h_1$). Upper fluid: silicon oil (depth at rest $h_2$). Dashed lines: interface and free surface at rest. A laser vibrometer locally measures the height of the interface, the water being dyed with white paint. Inset: Typical evolution of the interfacial wave height $\eta_I(t)$ as a function of time. $h_2=0.9~\rm{cm}$.}
\label{Issenmann_Setup}
\end{figure}

%In this paper, we study gravity-capillary wave turbulence at the interface of a two-layer stratified fluid. We show how the two propagative wave modes (in-phase or antiphase deformations of the two interfaces) become uncoupled when the upper fluid depth is increased, and how this decoupling influences the observed wave turbulence spectra. When the upper fluid becomes deep enough, the two modes are then completely uncoupled, one propagating at the interface and the other one at the free surface. A purely capillary wave turbulence spectrum is then observed on two frequency decades, whose exponent is in good agreement with weak turbulence theory. The article is organized as follows. We will first describe the experimental setup, then the experimental results and the model ones, before a comparison between both results. 

In this paper, we study gravity-capillary wave turbulence on the interface of a two-layer fluid with free upper surface.
Waves propagate both at the interface and at the free surface (either in phase or in antiphase), and this coupling depends strongly on the upper fluid depth. When this depth increases, we show that these two modes become progressively uncoupled explaining thus most of the observations on the wave turbulence spectra. When the upper fluid is deep enough, the two modes are then fully uncoupled, leading to the observation of a spectrum of purely capillary interfacial wave turbulence on two decades in frequency, fluids being of almost same density. The article is organized as follows. We will first describe the experimental setup, then the experimental results and the model, before comparing them to each other.

\section{Experimental setup}
The experimental setup is sketched in fig.~\ref{Issenmann_Setup}. Two fluids are placed in a circular 22 cm diameter plastic vessel. The lower fluid is water, and the upper fluid is a silicon oil (PDMS Dow Corning 200) \cite{PDMSDataSheet}. Their depths at rest are respectively $h_1$ and $h_2$. $h_1=4.3$ cm is fixed whereas $h_2$ is varied between $0$ and $0.9$ cm, thus $0\leq h_2/h_1 < 21\%$. Their kinematic viscosities are respectively $\nu_1=10^{-6}$~m$^{2}$/s and $\nu_2=10^{-5}$~m$^{2}$/s. Their densities are respectively $\rho_1=1000$ kg/m$^3$ and $\rho_2=935$ kg/m$^3$ \cite{ATC} leading to a small Atwood number $A=(\rho_1-\rho_2)/(\rho_1+\rho_2)=0.033$. The surface tension of silicon oil/air is $\gamma_S=20~\rm{mN/m}$  \cite{Issenmann2011}. The interfacial tension between water and silicon oil is $\gamma_I=25~\rm{mN/m}$\cite{Zhou2013}.

\begin{figure}[t]
\onefigure[width=8cm]{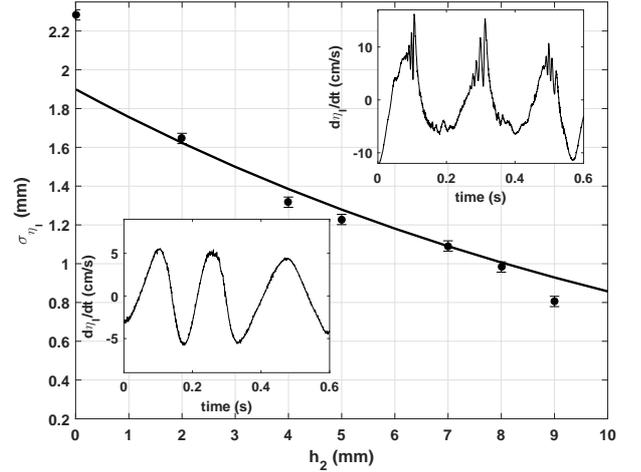}
\caption{Standard deviation of the interface wave height versus the upper fluid depth at rest, $h_2$. Solid line is from Eq.~(\ref{Issenmann_RapportAmplitudes_Equation}) with $f=4.5$ Hz, $\eta_S=1.9$ mm, mode~$+$. Top inset: Typical vertical velocity of surface waves, $h_2=0$. Bottom inset: Same for interface waves, $h_2=2$ mm. Wave steepness: 0.08.}
\label{Issenmann_Height}
\end{figure}

An electromagnetic shaker (LDS V406/PA 100E) vibrates horizontally a plexiglas blade that generates gravity-capillary waves at the interface between both fluids and at the free surface. The immersed part of the blade is fixed to $2$ cm regardless of $h_2$. The shaker is driven with a random forcing in amplitude and frequency between 1 Hz and 6 Hz. The rms amplitude and velocity of the blade is fixed to respectively 5 mm and 5 cm/s, regardless of the experiment presented here. A home made velocity sensor \cite{Falcon2008} is fixed to the shaker axis to measure the instantaneous blade velocity $V(t)$. The interface wave steepness ranges from 0.11 to 0.04 when $h_2$ is decreased.

A laser Doppler vibrometer (Polytec OFV506) placed above the setup measures the vertical velocity of the interface deformation at one point given by the position of the vertical laser beam (see fig.~\ref{Issenmann_Setup}). To wit, a white liquid dye ($\chem{TiO_2}$ particles plus a binding agent) \cite{dye} is added to the water bulk to make it slightly diffusing (typically $1\%$ in volume). The surface tension of dyed water/air is $\gamma_w=32$~mN/m \cite{Herbert10}. The velocity is extracted from the interference between the incident beam and the light back scattered by the diffusing fluid. After temporal integration, one thus obtains the interface height $\eta_I(t)$. The laser Doppler vibrometer has a sensitivity of order of 10$\mu$m. The free surface motions being dynamics, the angle of refraction of the laser within the upper fluid will change with respect to a free surface at rest. For our weak surface wave steepnesses ($< 0.13$), these changes correspond to horizontal displacements of the laser on the diffuse interface less than the beam diameter on the interface ($0.5$ mm). To avoid direct transmitted vibration from the shaker to the vibrometer, the latter is mechanically uncoupled from the shaker. Signals are then high-pass filtered ($> 0.5~\rm{Hz}$) to avoid possible residual low-frequency vibrations. They are acquired for $T=5$ minutes (or $30$ minutes to converge statistics to compute the probability density function (PDF) of the wave height). Surface tensions involved here are of close values: oil/air $\gamma_S=20$ mN/m \cite{Issenmann2011}, water/oil $\gamma_I=25$ mN/m \cite{Zhou2013}, and dyed water/air $\gamma_w=32$ mN/m \cite{Herbert10}. No significative change is thus expected in the model below when using water/oil interfacial tension instead of the unknown dyed water/oil one.

%{\color{red} Note that the laser refraction within the upper fluid was taken into account by dividing the data, obtained at the interface, by the refraction index of the upper fluid ($n=1.4$ \cite{PDMSDataSheet})

%Note that the white dye added to water certainly modifies its surface tension $\gamma_w$ and its viscosity but their orders of magnitudes remain valid. We will thus use their above values in the model below. 
%

\section{Experimental results}
\begin{figure}[t]
\onefigure[width=8cm]{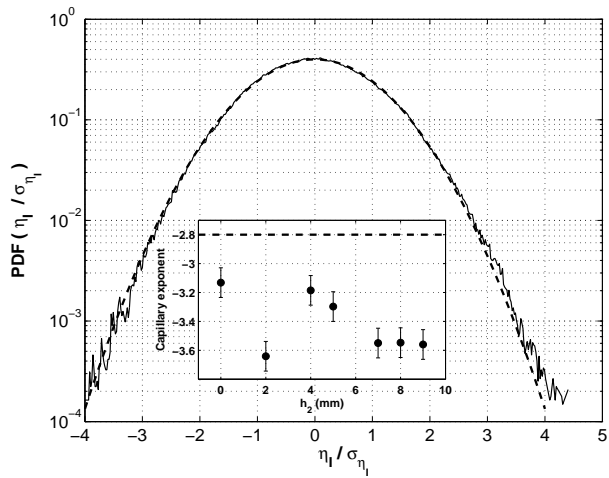}
\caption{Solid line: Probability density function (PDF) of the rescaled interfacial wave height, $\eta_I/\sigma_{\eta_I}$. Dashed line: Gaussian with zero mean and unit standard deviation. $h_2=4~\rm{mm}$. $\sigma_{\eta_I}=1.32~\rm{mm}$. Inset: Frequency-power law exponents of the capillary spectrum as a function of $h_2$. Dashed line: Theoretical value $-17/6$ for capillary wave turbulence.}
\label{Issenmann_PDF}
\end{figure}

A temporal recording of the interfacial wave height $\eta_I(t)$ is shown in the inset of fig.~\ref{Issenmann_Setup}. It displays an erratic behavior in response to the stochastic forcing. As shown in fig.~\ref{Issenmann_Height}, the standard deviation of the temporal recording of the interface wave height $\sigma_{\eta_I}\equiv\sqrt{\langle \eta_I^2(t)\rangle}$ is found to decrease strongly when the depth of the upper fluid, $h_2$, is increased. Temporal average is denoted by $\langle \cdot \rangle$. This experimental decrease is well described theoretically by the model presented below (solid line in fig.~\ref{Issenmann_Height}). We observe no perforation of the top layer regardless its depth. No break slope on the vertical velocity of the interfacial waves occurs that would be related on a change of index of refraction if perforation occurred. The vertical velocity is shown in the insets of Fig.~\ref{Issenmann_Height}. For a single fluid ($h_2=0$), capillary wave generations are observed near the crests of steep gravity-capillary waves whereas this effect is absent for a two-layer fluid ($h_2\neq 0$). The probability density function of the rescaled interface wave height, $\eta_I/\sigma_{\eta_I}$, is closed to a Gaussian (see fig.~\ref{Issenmann_PDF}), and is found independent of $h_2$ (not shown).

% The interface wave steepness~$\epsilon_I$ ranges from 0.11 to 0.08 when $h_2$ decreased. $\epsilon_I \equiv \sigma_{\eta_I}*k_m$ with $k_m$ the wavenumber corresponding to the maximum amplitude of the wave spectrum (occurring for a frequency $f_m=3.5$ Hz) located in the forcing range (see below).

 %\revision{Note that a slight asymmetry is found for a stronger forcing}, due to wave nonlinearity, as previously observed for free surface waves \cite{Falcon2007,Issenmann2013}. 

%Assuming that the distribution of free surface wave height, $\eta_S$, is also Gaussian, we infer that the PDF of their height difference, $\eta_I-\eta_S$, is also close to a Gaussian of standard deviation $\sigma_{diff}=\sqrt{\sigma_{\eta_I}^2+\sigma_{\eta_S}^2}$. We can then easily compute the probability to have $\eta_I-\eta_S$ larger than $h_2$, that quantifies the probability for  the lower fluid to emerge from the upper fluid layer. Since the prescribed rms velocity of the blade is kept constant for all the experiments, we assume $\sigma_{\eta_S}(h_2)$ to be constant regardless of $h_2$, and equal to its value measured for $h_2=0$, i.e. $\sigma_{\eta_S}(h_2)=2.3$ mm, both fluids being incompressible and viscous effects negligible at large scales (see below). Those probabilities are displayed in the inset of fig.~\ref{Issenmann_Height}, and show that lower fluid rarely emerges from the upper fluid layer, for most of our experiments. It is thus relevant to consider a continuous upper fluid layer with no interfacial crest emerging from the upper layer. Note that this is not true for the smallest depth used, $h_2=2~\rm{mm}$ since this probability reaches $24\%$. %{\it This effect might be the cause of the a{\color{red} b}normally small capillary {\color{red} wave turbulence} exponent observed for $h_2=2~\rm{mm}$ (see {\color{red} below, and} inset of fig. {\color{red} \ref{Issenmann_PDF}})}.

\begin{figure}
\onefigure[width=8.8cm]{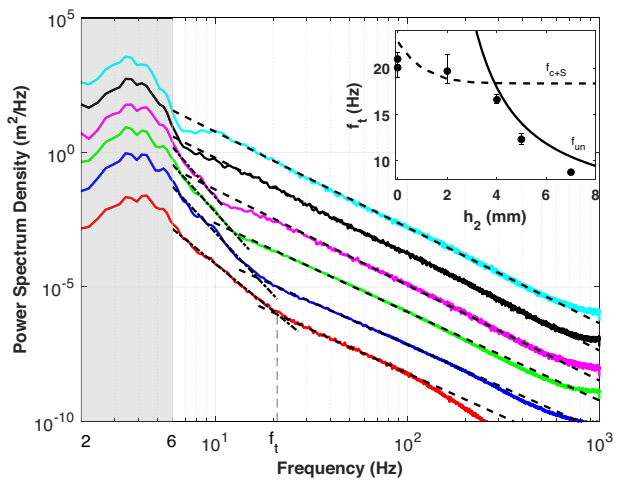}
\caption{(color online) Power spectrum density of the interface wave height, $S_{\eta_I}(f)$, for increasing upper fluid depths $h_2=0$, $4$, $5$, $7$, $8$ and $9~\rm{mm}$ (from bottom to top, shifted vertically for clarity). The dashed (resp. dash-dotted) lines are the best power-law fits of the capillary (resp. gravity) wave turbulence regime. Forcing frequencies $\leq 6$ Hz (represented by shaded area). Inset: Experimental crossover frequency $f_t$ between gravity and capillary regimes versus $h_2$ (symbols). Dashed line: theoretical crossover frequency $f_{c+S}$ (see text). Solid line: frequency $f_{{\rm un}}$ for which the interface and free surface waves of the mode $+$ become uncoupled (i.e. $\eta_I/\eta_S=1/10$ - see fig.~\ref{Issenmann_RapportAmplitude}).}
\label{Issenmann_Spectres}
\end{figure}

From the temporal recording of the interface wave height, $\eta_I(t)$ (see inset of fig.~\ref{Issenmann_Setup}), one computes its power spectrum density as the square modulus of the Fourier transform of $\eta_I(t)$ over a duration $T$: $S_{\eta_I}(f)\equiv |\int_0^T \eta_I(t)e^{i\omega t}dt|^2/(2\pi T)$, where $\omega=2\pi f$. Figure~\ref{Issenmann_Spectres} shows the power spectra of the interface wave height when the upper fluid depth increases. For $h_2=0~\rm{mm}$ (bottom curve), the usual gravity-capillary wave turbulence spectrum is observed as previously found in several recent studies \cite{Falcon2007,Denissenko2007,Cobelli2011,Issenmann2013,Deike2015}. Up to a cutoff frequency at  $\approx 100~\rm{Hz}$, related to dissipation, this spectrum is consistent with two different power-law regimes for frequencies above $f_t\approx 20~\rm{Hz}$ (capillary regime) and below $f_t$ (gravity regime) as expected by the weak turbulence theory \cite{Zakharovbook,Nazarenkobook} 
and already observed experimentally \cite{Falcon2007,Denissenko2007,Cobelli2011,Issenmann2013,Deike2015}. Note that the exponent of the capillary regime is slightly lower than its predicted value $-17/6$ (inset of fig.~\ref{Issenmann_PDF}). The crossover frequency $f_t$ between gravity and capillary wave turbulent regimes is linked to the capillary length $l_{cw}=\sqrt{\gamma_w/\rho_1 g}$ \cite{Falcon2007}. When $h_2$ is increased, the capillary regime is found to hold down to lower and 
lower frequencies (see curves from $h_2=4~\rm{to}~7~\rm{mm}$ in fig.~\ref{Issenmann_Spectres}), until no transition is clearly visible before reaching the forcing frequencies (see curves for $h_2\ge 8~\rm{mm}$). This crossover frequency $f_t$ is found to decrease up to a factor 2.5 (from 20 to 8 Hz, see inset of fig.~\ref{Issenmann_Spectres}) when $h_2$ is increased. Moreover, the cut-off frequency increases of a factor 7 (roughly from 100 to 700 Hz) when $h_2$ is increased. This leads to a significative extension of the inertial range of the capillary spectrum by more than one order of magnitude. A frequency power-law spectrum is then clearly observed on almost two decades. In the next part, we will explain the widening of the spectrum inertial range and the dependence of the crossover frequency on the upper fluid depth.

\section{Theoretical description}
\begin{figure}
\onefigure[width=8cm]{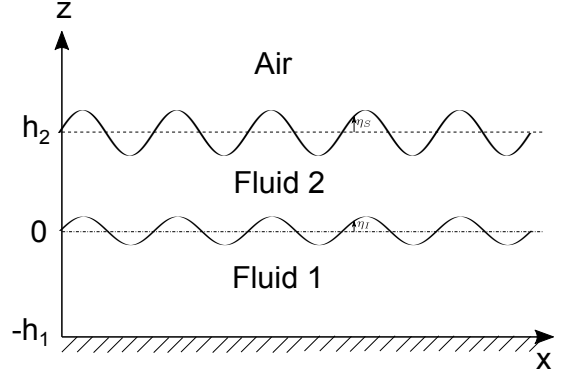}
\caption{Sinusoidal gravity-capillary waves propagating at the interface and at the free surface of two fluids of finite depths ($h_1$ and $h_2$ at rest). The upper surface is free. Case of in-phase deformations (mode $+$).}
\label{Issenmann_SchemaTheorie}
\end{figure}

To interprete our experimental results, we consider gravity-capillary waves propagating at both the interface and the free surface of two immiscible fluids of finite depths, the upper surface being free (fig.~\ref{Issenmann_SchemaTheorie}). The lower fluid is assumed to never emerge from the upper one. The dispersion relation of theses waves can be found in textbooks \cite{Landau1987,Lamb} but only when the capillary effects are neglected. To our knowledge, the theoretical derivation of the dispersion relation taking into account both the gravity and capillary effects for a two-layer fluid of finite depths with free upper surface were obtained only recently \cite{Woolfenden2011,Mohapatra2011}, as well as the one taking also into account the fluid viscosity effects \cite{Pototsky2016}. In the following, we will use the inviscid dispersion relation. The experimental validity of this hypothesis will be checked {\it a posteriori} (see below). 

Let a sinusoidal wave propagate along the $x$-axis with angular frequency $\omega$ and wave vector $k$ at the interface between both fluids, noticed $1$ and $2$, of finite depths (fig.~\ref{Issenmann_SchemaTheorie}). Fluid 1 is limited at the bottom by a rigid wall and fluid 2 is free at its surface. Experimentally, one has $6 \leq kh_1 \leq 500$ and $0.2\leq kh_2\leq 103$. Let $\eta_I$ be the wave height at the interface and $\eta_S$ be the wave height at the free surface. The system is assumed invariant along the $y$-axis, and the flows incompressible and inviscid. The interface at rest is located at $z=0$. The corresponding dispersion relation reads~\cite{Mohapatra2011,Woolfenden2011}

\begin{equation}
  a\omega^4+b\omega^2+c=0
  \label{Issenmann_RD}
\end{equation}
with
\begin{equation}
  \left \{
  \begin{array}{c c l}
      a & = & \rho_2\left[\rho_1+\rho_2\tanh(kh_1)\tanh(kh_2)\right]	\\
      b & = & -\rho_1(\rho_2gk+\gamma_Sk^3)\tanh(kh_2)+\\
	&   &  \rho_2\left(-\rho_1gk-(\gamma_I+\gamma_S)k^3\right)\tanh(kh_1)	\\	
      c & = & \left[\gamma_I\gamma_Sk^6+g\left(\rho_2(\gamma_I-\gamma_S)+\rho_1\gamma_S\right)k^4\right.	\\
	&   &	\left.-\rho_2(\rho_2-\rho_1)g^2k^2\right]\tanh(kh_1)\tanh(kh_2)	
  \end{array}
  \right.
  \label{Issenmann_RD_Explicite}
\end{equation}

Note that for $h_1=h_2=\infty$, Eqs.~(\ref{Issenmann_RD}) and (\ref{Issenmann_RD_Explicite}) are well reduced to the usual gravity-capillary dispersion relation between two infinite fluids with no free surface \cite{Lamb}

\begin{equation}
 \omega^2(k)=\frac{\rho_1-\rho_2}{\rho_1+\rho_2}gk+\frac{\gamma_I k^3}{\rho_1+\rho_2}\ {\rm .}
 \label{Issenmann_RD_FluidesInfinis}
\end{equation}

For $\rho_2=0$ and $h_2=\infty$, Eqs.\ (\ref{Issenmann_RD}) and (\ref{Issenmann_RD_Explicite}) lead to the usual gravity-capillary dispersion relation at the free surface of single fluid of finite depth \cite{Landau1987}

\begin{equation}
 \omega^2(k)=\tanh(kh_1)(gk+\gamma_I k^3/\rho_1)\ {\rm .}
 \label{Issenmann_RD_1Fluide}
\end{equation}

\begin{figure}
\onefigure[width=8cm]{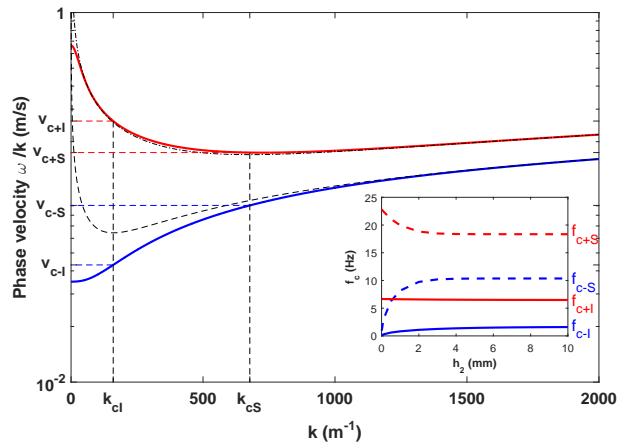}
\caption{(color online) Theoretical dispersion relation $\omega/k$ vs. $k$ for gravity-capillary waves in a two-layer fluid of finite depths from Eqs. (\ref{Issenmann_RD}) and (\ref{Issenmann_RD_Explicite}). Red line: in-phase mode (mode $+$). Blue line: anti-phase mode (mode $-$). $h_2=2~\rm{mm}$. Black dashed line: Dispersion relation at the interface between the two same fluids but for infinite depths (Eq.~\ref{Issenmann_RD_FluidesInfinis}). Black dash-dotted line: Dispersion relation at the free surface of a single fluid (2) of infinite depth (Eq.~\ref{Issenmann_RD_1Fluide} with $h_1=\infty$, replacing $\rho_1$ and $\gamma_I$ by $\rho_2$ and $\gamma_S$ respectively). Inset: Solid (resp. dashed) line: Crossover frequencies between capillary and gravity wave regimes at the interface (resp. at the free surface) for both modes as a function of $h_2$. Mode $+$ (red). Mode $-$ (blue).} %From top to bottom (in the large $h_2$ limit): $f_{c+S}$, $f_{c-S}$, $f_{c+I}$ and $f_{c-I}$.
\label{Issenmann_RelationDispersion}
\end{figure}

The dispersion relation in the general case, Eqs. (\ref{Issenmann_RD}) and (\ref{Issenmann_RD_Explicite}), have 4 solutions $\omega^2(k)=\frac{-b\pm\sqrt{b^2-4ac}}{2a}$ among which only 2 are real, that are plotted in solid lines (blue and red) in fig.~\ref{Issenmann_RelationDispersion}. We will call these solutions ``mode $+$'' and ``mode $-$'' according to the sign in the above expression of $\omega^2(k)$. Note that the mode $+$ has a higher phase velocity than the mode $-$ (semilog-y plot in fig.~\ref{Issenmann_RelationDispersion}). Figure~\ref{Issenmann_RelationDispersion} also shows that even for a thin enough upper fluid layer (as small as $2~\rm{mm}$), the wave dispersion relation is constituted of two branches corresponding to the two propagating modes. This differs strongly from the dispersion relation for interfacial waves between two infinite fluids (see dashed line) or for surface waves on the surface a single infinite fluid (see dash-dotted line). The interfacial and free surface waves are indeed not independent but are coupled to each other by those two propagative modes. Both modes propagate at both interface and free surface. The ratio between the wave heights at the free surface (S) and the interface (I) reads~\cite{Mohapatra2011}
\begin{multline}
\frac{\eta_S}{\eta_I}=\frac{\sinh(kh_2)}{\rho_2}\times \bigg[\rho_2\coth(kh_2)+\rho_1\coth(kh_1)\\ 
-\frac{[(\rho_1-\rho_2)g+\gamma_Ik^2]k}{\omega^2(k)}\bigg]\ {\rm .}
	\label{Issenmann_RapportAmplitudes_Equation}
\end{multline}

If $\eta_S/\eta_I>0$, the waves at the interface and the ones at the free surface propagate in phase, corresponding to the mode $+$ (also called barotropic \cite{Pucci2013} or zigzag \cite{Pototsky2016} mode) as illustrated in Fig.\ \ref{Issenmann_SchemaTheorie}. For $\eta_S / \eta_I <0$, they propagate in antiphase corresponding to the mode $-$ (also called varicose mode \cite{Pototsky2016}).

%The phase shift between the waves propagating at the interface and the ones at the free surface is $\Delta \Phi=\arctan[{\rm Im}(\eta_S/\eta_I)/{\rm Re}(\eta_S/\eta_I)]$ \cite{Pototsky2016}. Thus, for $\Delta \Phi=0$,} 
%\begin{equation}
%\begin{array}{l}
%{\displaystyle {\frac{\eta_S}{\eta_I}=\frac{\sinh(kh_2)}{\rho_2}\times}} \cdots \\ 
%\\
%		 \left[\rho_2\coth(kh_2)+\rho_1\coth(kh_1) -\frac{[(\rho_1-\rho_2)g+\gamma_Ik^2]k}{\omega^2(k)}\right]\ {\rm .}
%\end{array}
%	\label{Issenmann_RapportAmplitudes_Equation}
%\end{equation}

The modulus of the wave height ratio $\left| \eta_I/\eta_S \right|$ is plotted in fig.~\ref{Issenmann_RapportAmplitude} as a function of the wave frequency for different depths $h_2$, for both modes. We first note that the surface wave height is higher than that of the interfacial wave height for the mode $+$, regardless $h_2$. On the other hand, for the mode $-$, the interfacial wave height is higher than that of the surface wave height, regardless $h_2$. Second, within our experimental inertial range ($f \geq 6$ Hz) and for small $h_2$, wave heights at the interface and at the free surface are found to be of the same order for both modes, and are thus coupled (see insets of fig.~\ref{Issenmann_RapportAmplitude}). For large $h_2$, waves are uncoupled since the interface wave height is much smaller (resp. much higher) than that of the surface wave height for mode $+$ (resp. mode $-$): waves can be thus considered to propagate only at the free surface for the mode $+$ and only at the interface for the mode~$-$. Note that this decoupling is as strong as the wave frequency is large. These features are shown in insets of fig.~\ref{Issenmann_RapportAmplitude} for a fixed frequency.  Experimentally, the paddle forces both interfaces in phase at large scales, thus favoring the in-phase mode. However, a mixing between the two eigenmodes occurs in practice due to the random feature of the forcing. % inducing some out-of phase wave components.
%{\it Finally, note that in the deep water case ($kh_1\gg 1$ and $kh_2\gg 1$), one would have had $\eta_I/\eta_S=\exp{(-kh_2)}$ (mode $+$) and $0$ (mode $-$) \cite{Mohapatra2011}}.

\section{Interpretation}
Let us introduce the typical capillary lengths of the interface, $l_{c_{I}}$, and of the free surface, $l_{c_{S}}$. The corresponding wavenumbers are $k_{c_{I}}\equiv1/l_{c_{I}}=\sqrt{(\rho_1-\rho_2)g/\gamma_I}$ and $k_{c_{S}}\equiv1/l_{c_{S}}=\sqrt{\rho_2g/\gamma_S}$ ($\lambda_{c_{I}}\equiv 2\pi/ k_{c_{I}}\simeq 3.9$ cm ; $\lambda_{c_{S}}\equiv 2\pi/k_{c_{S}}\simeq0.9$ cm). Using the dispersion relation $\omega(k)$, those typical lengths correspond to 2 frequencies per mode: $f_{c-I}$, $f_{c-S}$ (mode $-$) and $f_{c+I}$, $f_{c+S}$ (mode $+$) as displayed in fig.~\ref{Issenmann_RelationDispersion} in the phase velocity space ($v \equiv \omega/k$ ; $k$) for a fixed depth $h_2$. Those frequencies correspond to the crossover frequencies between gravity and capillary wave regimes either at the interface (I) or at the free surface (S) for both modes ($+$ or $-$). The evolutions of these 4 frequencies with the upper fluid depth, $h_2$, are plotted in the inset of fig.~\ref{Issenmann_RelationDispersion}. They are almost independent of $h_2$ within our experimental range ($h_2\ge 2~\rm{mm}$) and only $f_{c+S}$ and $f_{c-S}$ lie within our experimental inertial range ($f\geq 6$ Hz). $f_{c+S}$ corresponds to the minimum of the upper branch of dispersion relation (see fig.~\ref{Issenmann_RelationDispersion}) except for small $h_2$. Moreover, the two modes are best studied when the densities of the two fluids differ slightly \cite{Mohapatra2011}, as in our study. Indeed, $l_{c_{S}}/l_{c_{I}}= \sqrt{(\gamma_s/\gamma_i)(\rho_1-\rho_2)/\rho_2}$. Thus, when the density difference decreases (or the interfacial tension increases), the gap between capillary lengths will increase, but does not modify qualitatively the results reported here.

\begin{figure}
\onefigure[width=8cm]{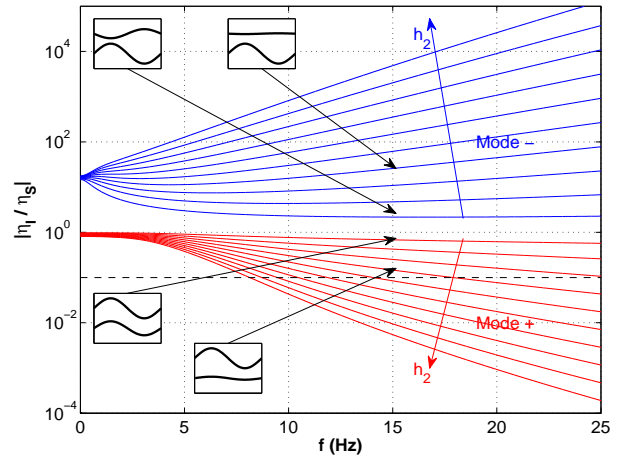}
\caption{(color online) Theoretical ratio from Eq.\ (\ref{Issenmann_RapportAmplitudes_Equation}) between the interfacial and surface wave heights as a function of their frequencies, for different upper fluid depth $h_2=1$ to $10~\rm{mm}$ with a step of $1~\rm{mm}$ (see arrows). Red lines: Mode $+$. Blue lines: Mode $-$. Dashed line: $\left|\eta_I\right|=\left|\frac{\eta_S}{10}\right|$. Insets show temporal evolutions of the surface and interfacial wave heights, for a fixed frequency $f=15$ Hz, $h_2=1$ or $4$ mm, and for both modes.}
\label{Issenmann_RapportAmplitude}
\end{figure}

%Moreover, the mode - propagates only at the interface then only $f_{c+S}$ is relevant in our problem.
%The mode - propagates in a purely capillary regime in our experimental frequency range.

Let us now interpret the experimental evolution of the crossover frequency $f_t$ between gravity and capillary wave turbulence regimes when $h_2$ is increased (see inset of fig.~\ref{Issenmann_Spectres}). As explained above, for small depths $h_2$, waves at the interface and at the free surface are coupled for both propagating modes $+$ or $-$. The theoretical crossover frequency $f_{c+S}$ at the free surface for the mode $+$ is found to well describe the data for small enough depths $h_2$ (see dashed line in the inset of fig.~\ref{Issenmann_Spectres}). When $h_2$ is increased, this is no longer the case. Indeed, as also explained above, when $h_2$ is increased, the waves propagating at the interface and at the free surface become progressively uncoupled for both modes: their relative height $\left| \eta_I/\eta_S \right|$ decreases strongly for the mode $+$ or strongly increases for mode $-$ (see fig.~\ref{Issenmann_RapportAmplitude}). We arbitrary decide that waves become uncoupled when $\left| \eta_I/\eta_S \right| \leq 1/10$ in the mode $+$, that is when the interface wave height becomes 10 times smaller than the free surface wave height. Interfacial waves can thus only propagate significantly on the other mode (mode $-$). This criterion corresponds to the dashed line in fig.~\ref{Issenmann_RapportAmplitude}. The intercepts of this dashed line and each solid line in fig.~\ref{Issenmann_RapportAmplitude}, for each depth $h_2$, thus give the frequencies $f_{{\rm un}}$ for which the interface and free surface waves of the mode $+$ become uncoupled. $f_{{\rm un}}$ is then found to decrease with $h_2$ in rough agreement with the experimental crossover frequency $f_t$ at large $h_2$ (see solid line in the inset of fig.~\ref{Issenmann_Spectres}). 

To sum up, for small fluid depths $h_2$, the interface waves (that we measure) propagate on both modes and are coupled with surface waves. The crossover frequency between the gravity and capillary wave turbulence regimes observed on the spectrum, $f_t$, is linked to the value of $f_{c+S}$ the capillary length at the free surface of the in-phase mode (mode $+$). The dependence of $f_t$ on $h_2$ is thus well described by $f_{c+S}(h_2)$ until the interfacial and surface waves decouple for large enough $h_2$, that is for $f_{{\rm un}} < f_{c+S}$ (see inset of fig.~\ref{Issenmann_Spectres}).

\section{Wave dissipation}
Let us now explain why the cut-off frequency of the wave spectrum increases strongly when an upper fluid is added. For a single fluid ($h_2=0$), there are three distinct mechanisms for the dissipation of wave energy: wave breaking, viscous damping and the generation of parasitic capillaries near steep crests of longer waves. Here, no wave breaking (i.e.~multi-valued interface) occurs since the wave steepnesses are small enough. Wave energy extraction by generation of high-frequency parasitic capillaries is indeed observed on the wavefront face as shown in the top inset of fig.~\ref{Issenmann_Height}, with a typical frequency ranging from 100 to 200 Hz. The latter comes from a resonance condition, $c_{gc}=c_c$, matching phase speeds between longer gravity-capillary waves ($c_{gc}$) and the shorter parasitic capillary waves ($c_c$)~\cite{Fedorov1998}. Using this condition and the dispersion relation of a single fluid, one finds a typical frequency of order of $200$ Hz for ripples propagating at the same phase speeds than the longer waves near the forcing scales ($f\sim 5$ Hz). This typical frequency gives also the typical dissipative scale which is found to be in good agreement with the frequency cut-off of the wave turbulence spectrum (see fig.~\ref{Issenmann_Spectres} for $h_2=0$). This dissipative scale depends theoretically slightly on the wave steepness \cite{Fedorov1998}. In presence of an upper layer of fluid ($h_2 \neq 0$), the resonance condition with the full dispersion relation [Eqs.~(\ref{Issenmann_RD}) and (\ref{Issenmann_RD_Explicite})] leads to a parasitic capillary frequency near 300 Hz (for mode $+$) regardless $h_2$ (no solution for the mode $-$).  However, the amplitude $\eta_I$ of this parasitic capillaries is predicted to be negligible ($\eta_I < 10$ $\mu$m) regardless $h_2$ by using Eq.\ (\ref{Issenmann_RapportAmplitudes_Equation}) with $f=300$ Hz and assuming $\eta_S\sim 1$ mm. Indeed, experimentally for a two-layer fluid, we observe no generation of parasitic capillaries near steep crest waves as shown in the bottom inset of fig.~\ref{Issenmann_Height}. Note that this absence of parasitic capillaries should not be confused with the effect of an oil monolayer on ocean calming gravity-capillary waves which is due to Marangoni dissipation~\cite{Alpers1989}. Thus, the main wave dissipation in our two-layer fluid system is the viscous dissipation occurring at higher frequency. %It will be used in the next Section.   %but also with other previous gravity-capillary wave turbulence studies, although not noticed \cite{}

\section{Time-scale separation} 
\begin{figure}
	\onefigure[width=8.8cm]{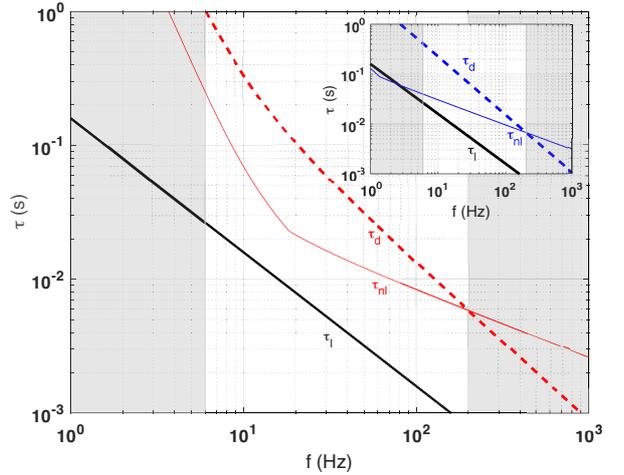}
	\caption{(color online) Typical time scales as a function of the wave frequency. Thick solid line: $\tau_l=1/\omega$. Thin solid line: $\tau_{nl}$. Dashed line: $\tau_d$. Mode $+$. $h_2=4$~mm. Shaded areas: forcing and dissipative scales. Inset: same for mode $-$.} %For both modes, the time-scale separations $\tau_l(f) \ll \tau_{nl}(f) \ll \tau_d(f)$ are fulfilled between the forcing and dissipative scales.
	%Black lines are computed for $h_2=0$.
	\label{Issenmann_Temps}
\end{figure}

Let us now consider the typical time scales involved in our experiment. Weak turbulence theory assumes a time-scale separation $\tau_l(f) \ll \tau_{nl}(f) \ll \tau_d(f)$, between the linear propagation time, $\tau_l$, the nonlinear interaction time, $\tau_{nl}$, and the dissipation time, $\tau_d$. The linear propagation time is $\tau_l=1/\omega(k)$. The dissipative time scale is linked to the viscous surface boundary layer, modeled by an inextensible infinitely thin film on the free surface, reading $\tau_d=\frac{2\sqrt{2}}{k(\omega)\sqrt{\nu\omega}}$ \cite{Lamb,Deike2012}. The kinematic viscosity of the two-layer fluid is $\nu=\frac{\mu_1\coth(kh_1)+\mu_2\coth(kh_2)}{\rho_1\coth(kh_1)+\rho_2\coth(kh_2)}$ with $\mu_1$ (resp. $\mu_2$) the dynamic viscosity of the fluid beneath (resp. above) the interface \cite{Kumar1994}. We show in fig.~\ref{Issenmann_Temps} that the condition $\tau_l(f)  \ll \tau_d(f)$ is well satisfied in our experimental frequency range for both modes. It thus validates {\it a posteriori} the use here of an inviscid dispersion relation. Finally, for wave turbulence to take place, the typical time scale $\tau_{nl}$ of nonlinear wave interactions has to satisfy $\tau_l(f) \ll \tau_{nl}(f) \ll \tau_d(f)$. For capillary waves, $\tau_{nl}=\xi_c k^{-3/4}$ \cite{Zakharov1967,Deike2013} while for gravity waves, $\tau_{nl}=\xi_gk^{-3/2}$ \cite{Deike2013}. Since the constants $\xi_c$ and $\xi_g$ are experimentally unknown, $\tau_{nl}$ is plotted in fig.~\ref{Issenmann_Temps} assuming $\tau_{nl}$ continuous at the experimental crossover frequency, $f_t$, between gravity and capillary regimes, and $\tau_{nl}(f_{\rm cut})=\tau_d(f_{\rm cut})$ at the cut-off frequency of the experimental capillary power-law spectrum ($f_{\rm cut} \simeq 200$ Hz for $h_2$=4 mm - see fig.~\ref{Issenmann_Spectres}). Figure~\ref{Issenmann_Temps} shows that the time-scale separation is valid on the whole inertial range of our experiment (between shaded areas). This time-scale separation is also valid regardless the upper fluid depth $h_2$.

%($f_{\rm cut}\simeq 100$ and 250 Hz for $h_2$=0 and 4 mm, respectively -

%The viscous time $\tau_d$ was calculated at the interface when considering the mode - since it propagates almost only at the interface. $\tau_d$ has to be calculated at the free surface when considering mode +, since the viscous time is lower at the free surface than at the interface.

\section{Conclusion}
We studied gravity-capillary wave turbulence on the interface between two immiscible fluids with free upper surface. Waves propagate both at the interface and at the free surface [either in phase (mode $+$) or in antiphase (mode $-$)], and this coupling depends strongly on the upper fluid depth. When this depth is increased, these two modes become progressively uncoupled, and we show that this decoupling explains most of the observations on the wave turbulence spectra. Indeed, the crossover frequency between the gravity and capillary wave turbulence regimes is experimentally found to decrease by more than a factor $2$ when the upper fluid depth is increased. At small depths, interfacial and surface waves are coupled, and this crossover is linked to the value of the capillary length at the free surface of the in-phase mode. At large enough depth, they become uncoupled when the interfacial wave heights of this mode become negligible with respect to the surface wave heights. Interfacial waves (that we measure) can thus only propagate significantly on the other mode. The crossover frequency is then well described by this decoupling criterion depending on the upper fluid depth. Finally, when the upper fluid is deep enough, the two modes are then fully uncoupled. This leads to the observation of a spectrum of purely capillary interfacial wave turbulence on two decades in frequency, fluids being of almost same density. This frequency range is comparable to what was previously observed at the interface between two immiscible deep fluids of almost equal densities with no upper free surface \cite{During2009} or during microgravity experiments with a single fluid \cite{Falcon2009}.

To our knowledge, we report the first experimental study of wave turbulence in a two-layer fluid system with free upper surface. The interface wave steepness is weak enough to reach a weak wave turbulence regime involving a linear coupling between the two propagative modes. The linear dispersion relation of this system being only derived recently, we validated it experimentally (from the depth-dependence of the crossover frequency between the gravity and capillary wave turbulence regimes). In the future, we plan to increase the wave steepness to study nonlinear coupling between modes in the wave turbulence context. To wit, a simultaneous measurement of the surface and interface elevations will be also performed. At last, this study may be useful to better understand nonlinear wave dynamics within a two-layer fluid in presence of surface and interfacial tensions such as oil spilling in oceanography and gravity-capillary solitary waves. The reported phenomenon is more general and should be shown up in other wave turbulence systems involving the coupling between surface and interfacial waves.  
\acknowledgments
B. I. thanks CNRS for funding him as a one year postdoctoral research fellow. This  work has been partially supported by ANR Turbulon 12-BS04-0005.

\end{document}